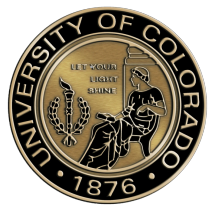

# Big Data, Big Decisions: Choosing the Right Database

Literature review by: **Mohamed Hassan**[*1]
*grad student*
*Dept. of computer science*
*Mohamed.hassan@colorado.edu*
*https://orcid.org/*
*https://scholar.google.com/*

[1*] College of Engineering and Applied Sciences, University of Colorado, Boulder, CO, United States of America

## I. KEYWORDS

Big data,
Relational Database "RDBMS",
Atomicity, Consistency, Isolation, and Durability "ACID",
Graph, Key-Value pairs,
Tables,
Normalization,
JSON,
XML Database,
XML,
SQL,

## II. ABSTRACT

In the burgeoning era of big data, selecting the optimal database solution has become a critical decision for organizations across every industry.

Big data[1] demands a powerful database solution. Traditionally, SQL Database[2] Database ruled, offering a structured approach familiar to many organizations. However, big data's complexity and unstructured nature challenge SQL Database's limitations. Enter NoSQL Database: flexible and scalable, making them ideal for big data's ever-changing nature.

We'll explore the key differences between SQL and NoSQL Database. Performance-wise, SQL Database shines for structured queries. Its standardized language (SQL)[3] ensures data consistency and complex analysis. But for big data's unstructured formats, this rigidity becomes a hurdle.

NoSQL[4] offers a welcome contrast. Its flexible schema allows for diverse data formats and evolving structures, perfect for undefined or frequently changing data models. Additionally, NoSQL boasts superior horizontal scalability, distributing data across multiple servers for cost-effective growth.

Understanding these key differentiators empowers organizations to choose the optimal database for their big data needs.

## III. INTRODUCTION

In the world of data management, the rapidly growing area of big data offers both intriguing possibilities and substantial obstacles. Big data, defined by its enormous size, complex structure, and varied sources - which include everything from social media engagement and sensor data to financial transactions and scientific measurements - is the gateway to deep insights across a wide range of fields. However, to effectively tap into this potential, it's crucial to strategically choose the right database technologies.

Historically, the realm of relational Database[5], controlled by the Structured Query Language (SQL), has been the foundation of data organization and retrieval. SQL Database are adept at systematically organizing data into well-defined tables with predetermined schemas. This orderly method makes them perfect for tasks involving structured data analysis, such as customer relationship management[6] or financial reporting. SQL, a potent query language specifically created for these Database, enables users to perform intricate analyses and manipulate data accurately. However, as big data's volume and complexity continue to rise, the inherent constraints of SQL Database become apparent. Their inflexible schema finds it difficult to adapt to big data's constantly changing nature, often leading to significant hurdles in storing and managing unstructured or semi-structured data, a common feature of many big data sources.

To overcome these limitations, NoSQL Database have surfaced as an attractive alternative, introducing a paradigm shift with their adaptable schema design. Unlike SQL Database, NoSQL Database adopt a more dynamic approach, easily accommodating unstructured, semi-structured, and even rapidly changing data formats. This inherent adaptability makes them especially skilled at handling big data applications that deal with continually changing information streams, such as social media sentiment analysis or sensor data from Internet of Things (IoT)[7] devices. Additionally, NoSQL Database offer the benefit of horizontal scalability[8], a robust feature that enables the easy addition of more servers to handle increasing data volumes. This is a stark contrast to SQL Database, which typically necessitate costly upgrades to a single server, often becoming an expensive bottleneck for big data management.

Therefore, the choice between SQL and NoSQL Database depends on the specific attributes of the data and the intended application. For situations requiring detailed analysis of well-structured data and the use of complex queries, SQL Database continue to be a reliable option. However, for the ever-evolving world of big data, characterized by large scale, varied formats, and continuous evolution, NoSQL Database provide an unmatched level of flexibility and adaptability, cementing their role as the champions of the big data age.

## IV. SQL: The Relational Database

SQL Database, also known as Relational Database (RDBMS), are renowned for their ability to execute Atomicity, Consistency, Isolation, and Durability ACID transactions[9] at high speeds. SQL Database's abstraction from data storage and indexing enables its uniform application across various problems and data set sizes, allowing it to operate efficiently across clustered replicated data stores. Structured Query Language (SQL) has been a dominant force for several decades and continues to receive substantial investment from big data companies and organizations such as Google, Facebook, Cloudera, and Apache.

However, relational Database face a range of limitations due to the continuous growth of stored and analyzed data. These limitations include restrictions on scalability and storage, efficiency loss in querying as data volume increases, and challenges in storing and managing larger Database. Companies like Google, Amazon, Facebook, and LinkedIn were among the first to identify these significant limitations of SQL database technology when supporting big data and large user requirements.

SQL is a programming language that enables users to query, manipulate, and alter data in a relational database. SQL Database, organized into columns and rows within a table, use a relational model that is most effective with well-defined structured data, such as names and quantities, where relationships exist between different entities. SQL Database can scale vertically, meaning the maximum load can be increased by adding more storage components like RAM or SSD. While this may mean that SQL Database are sometimes limited by server resources, cloud-based storage and other technologies can enhance SQL Database scalability.

SQL Database can read large data sets quickly compared to NoSQL, making it useful in scenarios where multiple data reading tasks need to be performed simultaneously. However, SQL Database requires a schema to be defined in Database before data storage and retrieval, which means SQL Database may take longer to prepare compared to NoSQL Database.

## V. NoSQL: The Non-Relational Database

NoSQL Database, also referred to as "Not Only SQL," is an alternative to SQL Database that doesn't necessitate fixed table schemas. NoSQL typically scales horizontally and avoids significant join operations on data. It can be described as structured storage, with relational Database as a subset. NoSQL Database encompasses a wide array of Database, each with a distinct data storage model. The most common types include Graph, Key-Value pairs[10], Columnar, and Document. NoSQL, a database technology propelled by Cloud Computing, the Web, Big Data, and Big Users, is now leading the way for popular internet companies such as LinkedIn, Google, Amazon, and Facebook to overcome the limitations of the 40-year-old RDBMS.

NoSQL Database are non-relational Database that store data differently than the tabular relations used in SQL Database. While SQL Database are optimal for structured data, NoSQL Database are suitable for structured, semi-structured, and unstructured data. NoSQL Database scale horizontally[8], meaning they use multiple nodes in a cluster to manage increased workloads, allowing data architects to easily scale them by adding more servers to clusters. NoSQL Database don't adhere to a rigid schema but instead have more flexible structures to accommodate their data types. Instead of using SQL to query the database, NoSQL Database use various query languages, with some not having a query language at all. With the rise in Web and mobile applications, emerging trends, shifting online consumer behavior, and new data classes, the industry's projects require a database technology that can provide a scalable, flexible solution to manage and access data. NoSQL technologies are the only solution available to effectively meet these needs.

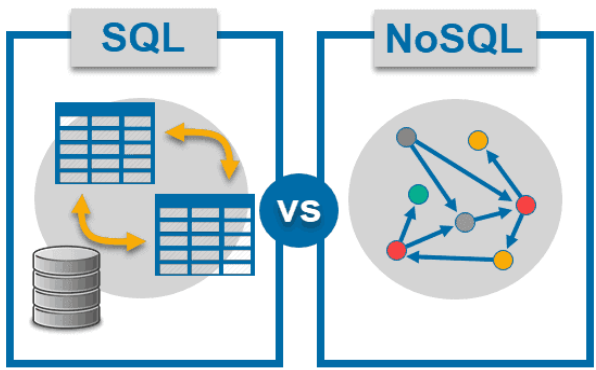

[Image 1]

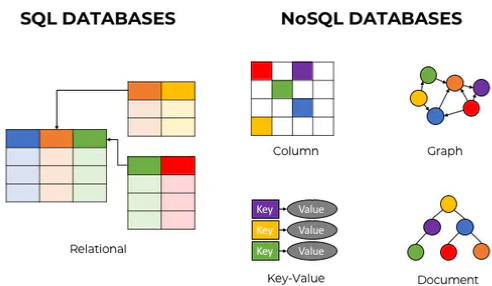

[Image 2]

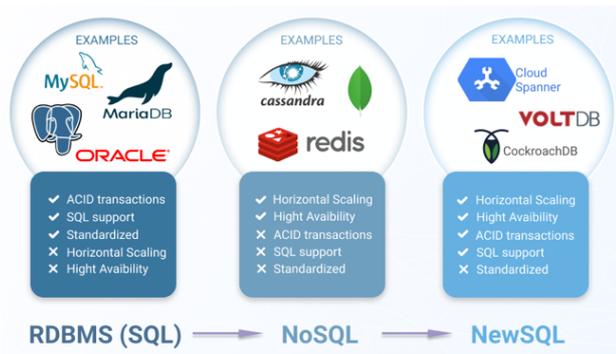

[Image 3]

## VI. COMPARING SQL AND NoSQL

### 1. Enables Interaction:

SQL is a declarative query language that facilitates interaction. On the other hand, NoSQL's MapReduce is a procedural query method. MapReduce not only requires users to know what they want but also how to obtain it. This distinction is crucial for two reasons.

Firstly, declarative SQL queries are simpler to construct, making database querying accessible to analysts, operators, managers, and others.

Secondly, separating the 'what' from the 'how' enables the database engine to use internal data to choose the most efficient algorithm. If the physical layout or indexing of the database changes, an optimal algorithm can still be computed. In a procedural system, however, a programmer must revisit and modify the original 'how,' which can be costly and prone to errors.

### 2. Speed:

SQL Database is a relational database that necessitates a high degree of Normalization[11]. This means that data must be divided into multiple tables[12] to Strengthen data reliability by ensuring its consistency and accuracy. While Normalization[11] contributes to efficient data management, the complexity of navigating through several related tables can slow down data processing in SQL-based relational Database.

In contrast, NoSQL Database like Couchbase, Cassandra, and MongoDB store data in flat collections. Here, data is often duplicated, and individual data pieces are rarely partitioned. Instead, they are stored as complete entities. This approach simplifies and accelerates read and write operations to individual entities.

### 3. Flexibility:

Relational and NoSQL data models have distinct characteristics. In the data model, relational data is divided into numerous interconnected tables consisting of rows and columns, with foreign keys in columns establishing the connections between tables.

To execute a query, information from multiple tables must be gathered and merged, a process that can involve hundreds of tables in modern enterprise applications.

Data writing also necessitates coordination across many tables. Relational Database are typically efficient for managing low-volume data flowing at a slow pace. However, contemporary applications often demand the rapid writing and reading of extensive data volumes, a requirement that NoSQL Database are designed to meet. Unlike relational Database, NoSQL or "NoREL" (non-relational) Database do not depend on interconnected tables to organize and store information.

### 4. For the type of data to be stored:

SQL Database aren't well-suited for hierarchical data storage. In contrast, NoSQL Database are more effective for this purpose as they use a key-value pair method for storing data, similar to JSON[13] data. NoSQL Database are often the preferred choice for handling large datasets, such as big data. Hbase is an example of this type of database.

### 5. Rapid Development:

NoSQL Database are typically less complicated and easier to deploy compared to SQL Database. Altering the way data is stored or modifying the queries in NoSQL Database is straightforward. Large-scale data modifications can be achieved through simple refactoring and batch processing, avoiding the need for complex migration scripts and downtime. It's even simpler to temporarily remove nodes from a cluster for modifications and reintegrate them, as the replication features ensure data synchronization and distribution of the new data design to other servers in the cluster.

### 6. Supports JSON:

A few years back, numerous SQL systems incorporated support for XML documents[14]. Presently, as JSON is becoming a favored data interchange format, SQL vendors are integrating support for JSON types. There are compelling reasons for supporting structured data types, considering the agile programming methodologies of today and the uptime demands of web-based infrastructures. Oracle 23c, PostgreSQL 16.2, VoltDB 13.2.1, among others, support JSON, often outperforming native JSON NoSQL Database stores in performance benchmarks. SQL Database is set to continue gaining market share and attracting new investments and implementations. NoSQL Database that offer proprietary query languages or basic key-value semantics without significant technical differentiation face challenges. Contemporary SQL Database systems often match or surpass their scalability while providing richer query semantics, having established and trained user bases, extensive ecosystem integration, and deep enterprise adoption.

Relational Database and NoSQL Database data models are very different. The relational model takes data and separates it into many interrelated tables that contain rows and columns. These tables reference each other through foreign keys that are stored in columns as well. When a user needs to run a query on a set of data, the desired information needs to be collected from many tables – often hundreds in today's enterprise applications – and combined before it can be provided to the application. Similarly, when writing data, the write needs to be coordinated and performed on many tables. When data is relatively low-volume, and when it is flowing into a database at a low velocity, a relational database is usually able to capture and store the information. But today's applications are often built on the expectation that massive volumes of data can be written (and read) at speeds near real-time. NoSQL Database have a very different model. At the core, NoSQL Database are really "NoREL," or non-relational, meaning they do not rely on tables and the links between tables in order to store and organize information."

### 7. Scalability:

One of the key advantages of NoSQL Database such as HBase for Hadoop, MongoDB, and Couchbase is their ability to easily scale horizontally across multiple servers or nodes to manage large amounts of data. For example, if you run an eCommerce website like Amazon and it becomes an instant hit, you'll have a massive influx of customers visiting your site. In such a scenario, if you're using a relational database like SQL, you would need to carefully replicate and restructure the database to meet the growing customer demand.

### 8. For DB types:

From a broad perspective, SQL Database can be categorized as either open-source or proprietary from commercial providers. NoSQL Database, on the other hand, can be classified according to their data storage method, such as graph Database, key-value store Database, document store Database, column store Database, and XML Database[15].

### 9. Data recovery:

Recovering data between SQL and NoSQL Database has some key differences due to their fundamental structures.

> Recovery Methods for SQL Database
> Transaction Logs, Backups, Point-in-Time Recovery (PITR)

> Recovery Methods for NoSQL Database
> Replication, Snapshots, Logging

In terms of data recovery, particularly during natural disasters in big data applications, NoSQL Database are easier to restore. As you may be aware, NoSQL Database is an unstructured database where data is stored in a document format.

### 10. Data security:

Compared to SQL database, NoSQL does not provide security as mature as SQL Database which is one of the main problems arise in big data application.

## VII. CONCLUSION

When managing substantial amounts of data, the decision between SQL and NoSQL hinges on the particular needs of the application.

By carefully evaluating these factors, organizations can illuminate the most suitable technology for navigating the evolving big data landscape.

SQL is appreciated for its data validity assurance and its efficiency in quickly reading large structured data sets. Conversely, NoSQL excels when rapid availability of big data is a priority. It's also a suitable option when a company anticipates scaling due to evolving requirements. In summary, both SQL and NoSQL possess their own advantages and disadvantages. The selection between the two should be dictated by the specific requirements of the application and the type of data involved.

In his paper titled 'No Silver Bullet', Fred Brooks made the following prediction/assertion "There is no single development, in either technology or management technique, which by itself promises even one order-of-magnitude improvement within a decade in productivity, in reliability, in simplicity."[16]

I hope that this Literature paper has improved your understanding of the benefits and drawbacks of both Traditionally SQL and NoSQL Database, enabling you to make an informed decision for your next project.

Images

## IX. ACKNOWLEDGMENTS

The author expresses his gratitude for the provision of a student account, which has enabled him to access the entire library of Oreilly.

## X. LLM DECLARATION

The author, being a non-native English speaker, utilized the GPT-4 model as a refinement instrument in the crafting of this Literature paper.

- Rectify grammatical errors and misspellings.
- Maintain linguistic uniformity and fluency.